\documentclass[floatfix,lengthcheck,showpacs,amssymb,amsmath,amsfonts,twocolumn]{aastex63}
\usepackage{CJK}

\usepackage[utf8]{inputenc}
\usepackage{color}
\usepackage{graphicx}
\graphicspath{{figs/}}

\usepackage{float}
\usepackage{subfigure}
\usepackage{amsmath}
\usepackage{afterpage}
\usepackage{multirow}
\usepackage{tikz}
\usetikzlibrary{arrows,shapes,trees,decorations.pathreplacing}
\usepackage{import}
\usepackage{svn-multi}
\usepackage{hyperref}
\usepackage{gensymb}
\usepackage{longtable}

\hypersetup{
    colorlinks=true,
    linkcolor=blue,
    filecolor=magenta,      
    urlcolor=cyan,
}

\usepackage{acro}

\DeclareAcronym{GR}{
	short = GR,
	long  = general relativity
	}
	
\DeclareAcronym{BH}{
	short = BH ,
	long  = black hole
}

\DeclareAcronym{BBH}{
	short = BBH ,
	long  = binary black hole
}
\DeclareAcronym{BNS}{
	short = BNS ,
	long  = binary neutron star
}
\DeclareAcronym{AGN}{
	short = AGN,
	long  = active galactic nuclei
}

\DeclareAcronym{GW}{
	short = GW ,
	long  = gravitational wave
}

\DeclareAcronym{CBC}{
	short = CBC,
	long  = compact binary coalescence
}

\DeclareAcronym{SNR}{
	short = SNR,
	long  = signal-to-noise ratio
}

\DeclareAcronym{PSD}{
	short = PSD,
	long  = power spectral density
}

\DeclareAcronym{GWTC}{
	short = GWTC,
	long  = gravitational wave transient catalog
}

\DeclareAcronym{IMR}{
	short = IMR,
	long  = inspiral-merger-ringdown
}

\DeclareAcronym{EOB}{
	short = EOB,
	long  = effective-one-body
}

\DeclareAcronym{FAR}{
	short = FAR,
	long  = false alarm rate
	}

\usepackage[capitalise]{cleveref}
\crefname{figure}{Fig.}{Figs.}
\Crefname{figure}{Fig.}{Figs.}

\def\be{\begin{equation}}
\def\ee{\end{equation}}
\def\({\left(}
\def\){\right)}
\def\[{\left[}
\def\]{\right]}

 \newcommand{\bqn}{\begin{eqnarray}}
 \newcommand{\eqn}{\end{eqnarray}}

 \newcommand{\f}{\frac}
 \newcommand{\p}{\partial}
\newcommand{\red}{\textcolor{black}}
\newcommand{\pycbc}{\texttt{PyCBC }}
\newcommand{\msun}{\ensuremath{\mathrm{M}_{\odot}}}
\newcommand{\etwo}{\ensuremath{e_\mathrm{20Hz}}}

\acsetup{patch/longtable=false}

\begin{document}
\begin{CJK*}{UTF8}{gbsn}
\title[]{Search for gravitational waves from eccentric binary black holes with an effective-one-body template}

\correspondingauthor{Yi-Fan Wang}
\email{yifan.wang@aei.mpg.de}

\author[0000-0002-2928-2916]{Yi-Fan Wang (王一帆)}
\affil{Max-Planck-Institut f{\"u}r Gravitationsphysik (Albert-Einstein-Institut), Am M{\"u}hlenberg 1, D-14476 Potsdam, Germany}

\author[0000-0002-1850-4587]{Alexander H. Nitz}
\affil{Department of Physics, Syracuse University, Syracuse, NY 13244, USA}

\keywords{gravitational waves ---  compact binaries --- black holes --- eccentricity}

\begin{abstract}
As gravitational wave astronomy has entered an era of routine detections, it becomes increasingly important to precisely measure the physical parameters of individual events and infer population properties. Eccentricity is a key observable, suggesting that binaries form in a dense stellar environment through dynamical encounters. This work performs the first matched-filtering search for gravitational waves from eccentric binary black holes (BBHs) covering the mass range $[5, 200]~M_\odot$ and eccentricity at 20 Hz up to 0.5 with a newly developed effective-one-body waveform model. Throughout the third observation run of LIGO, Virgo, and KAGRA, we identify 28 BBH events with a false alarm rate below once per 100 yr; all of which were previously reported in the GWTC-3 and 4-OGC catalogs. Additional candidates with false alarm rates between once per 1 and 100 yr are also reported. We perform an injection campaign to characterize the sensitive volume time of our search pipeline. Assuming that none of the eccentric BBH events were missed by previous searches, our results provide constraints on the event rate of eccentric BBHs in the mass range [5, 30] $M_\odot$. For a 30-30 $M_\odot$ BBH with eccentricity 0.5, the event rate is limited to less than 0.06 Gpc$^{-3}$ yr$^{-1}$; this marks an order of magnitude improvement for sensitive volume compared with the previous search with a minimally modeled algorithm without using templates.
\end{abstract}
\acresetall

\section{Introduction} \label{sec:intro}

The observation of \acp{GW} has opened up a novel avenue to investigate the characteristics of compact binary mergers.
To date, LIGO \citep{TheLIGOScientific:2014jea}, Virgo \citep{TheVirgo:2014hva}, and KAGRA \citep{Aso:2013eba} (LVK) collaboration has announced around 100 \ac{GW} events during their three observation runs \citep{GWTC-1, GWTC-2, KAGRA:2021vkt}. 
From the ongoing fourth observation, more than 200 additional event candidates have been identified through the low-latency online searches.
These \ac{GW} events enable parameter estimation for individual \acp{BBH} or neutron stars, as well as their property as a population.
The inferred event rate, mass, and spin distribution of the source population are key observables to investigate the formation mechanism of binary progenitors and to offer insights into stellar physics \citep{KAGRA:2021duu}.

Among the observables of \ac{GW} events, eccentricity is a critical evidence suggesting that the binary is formed through dynamical interaction within a dense stellar environment.
In contrast, isolated binaries would have been circularized before entering the sensitive frequency band of LIGO and Virgo due to \ac{GW} emission \citep{peter1,peter2}.
Astrophysical models have predicted that eccentric \acp{BBH} can originate within hierarchical triple systems located in globular clusters through the Kozai-Lidov mechanism \citep{Wen:2002km,Antonini:2017ash}, or arise in \ac{AGN} disks as a result of dynamic interactions with a dense population of stars and gas \citep{Samsing:2020tda,Tagawa:2020jnc}, among other scenarios.
Eccentric \acp{BBH} might also come from more speculative origins from the primordial Universe, where they could have eccentricity through dynamical encounters in the late Universe \citep{Wang:2021qsu}.

So far, all events in the \ac{GWTC} are detected by matched-filtering searches with a quasi-circular waveform template bank \citep{Roy:2017oul,Roy:2017qgg}.
A fraction of the \acp{BBH}, typically those with large mass, are also detected by the minimally modeled search without using a waveform template \citep{Drago:2020kic}.
Nevertheless, efforts to find eccentric binaries are ongoing.
LVK collaboration has performed a targeted search on eccentric \ac{BBH} with a minimally modeled coherent excess power method \citep{LIGOScientific:2019dag, LIGOScientific:2023lpe}.
The non-detection in the third observational run has been used to constrain the event rate upper limit on eccentric binaries with total mass $\geq 70 M_\odot$. 
On the other hand, matched-filtering searches are performed to find eccentric binary neutron stars and neutron star-black holes using an inspiral-only post-Newtonian waveforms \citep{Nitz:2019spj, Dhurkunde:2023qoe}.
Another matched-filtering search was conducted by Ref.~\citep{Nitz:2021mzz, Nitz:2022ltl} focusing on eccentric binaries with at least one subsolar mass, again using an inspiral-only waveform template.

Although the standard searches employ a template bank with quasi-circular waveforms, the method maintains some sensitivity to detecting binaries in eccentric orbits \citep{Ramos-Buades:2020eju}.
Several evidences of eccentricity in the current detection catalog have been reported.
\cite{Romero-Shaw:2020thy,Romero-Shaw:2021ual,Romero-Shaw:2022xko,Romero-Shaw:2025vbc} have reported in total four \ac{GW} events, GW190521, GW190620, GW191109, GW200208\_222617 with significant support for eccentricity $>$ 0.1 at \ac{GW} frequency of 10 Hz.
\cite{Gupte:2024jfe} has reported three \ac{BBH} events, GW200129, GW190701, and GW200208\_222617 with non-negligible eccentricity by the standard of a log$_{10}$ Bayes factor greater than $1$ comparing a spin-aligned, eccentric waveform with a quasi-circular spin-aligned, or spin-precessing model.
GW190521, an exceptional \ac{BBH} with large total mass $\sim150\msun$ \citep{LIGOScientific:2020iuh} was reported to have eccentricity using a spin-aligned eccentric waveform \citep{Romero-Shaw:2020thy} or using numerical relativity simulations as templates \citep{Gayathri:2020coq} (However, more recent studies \cite{Gamboa:2024hli, Gupte:2024jfe, Ramos-Buades:2023yhy} have not found evidence of eccentricity in GW190521).
Recently, with the development of a post-Newtonian spin-precessing eccentric waveform model, \cite{Morras:2025xfu} reports evidence of eccentricity in a neutron star - black hole binary GW200105, with eccentricity $\sim 0.145$ at 20 Hz using a uniform prior on eccentricity, excluding zero at more than 99\% confidence.
However, the evidence is greatly reduced using a log-uniform prior. 
Parameter estimation on the two binary neutron star mergers, GW170817 and GW190425, has also been performed, but no evidence for eccentricity was shown \citep{Lenon:2020oza}.
Active investigations are in progress to look for evidence of eccentricity within the \ac{GW} event catalog.

Without a targeted search with an eccentric waveform model, some real events with non-negligible eccentricity might be missed in the first place \citep{Gadre:2024ndy}, or their statistical significance would be underestimated due to the mismatch against a quasi-circular template bank from signal-consistency tests \citep{Allen:2004gu}.
Incorporating eccentricity is also essential for accurately determining the source properties; neglecting it may lead to errors in parameter estimation \citep{Divyajyoti:2023rht} and tests of general relativity \citep{Narayan:2023vhm,Shaikh:2024wyn, Roy:2025xih}.
Even when eccentric binaries are not detected, a waveform approximant model or numerical relativity waveforms \citep{Bhaumik:2024cec}, remain necessary to robustly measure the sensitivity of a search pipeline towards a population of sources to convert the search results to a constraint on the astrophysical event rate.
With a new effective-one-body \ac{BBH} waveform that incorporates eccentricity, we report the first match-filtering search for \ac{GW} from eccentric \ac{BBH} in this work.
We present the waveform model and search parameter space in section \ref{sec:parameter}, introduce the search strategy in section \ref{sec:strategy}, discuss the results in section \ref{sec:results} and constrain the astrophysical event rate density in section \ref{sec:astro}.
We conclude and point out possible future developments in section \ref{sec:conclusion}.

\section{Search parameter space}
\label{sec:parameter}
There has been a rapid progress recently in the development of \ac{IMR} waveform model accounting for eccentric spin-aligned binaries \citep{Huerta:2016rwp,Huerta:2017kez,Hinder:2017sxy,Islam:2024tcs,Ramos-Buades:2019uvh,Chattaraj:2022tay,Manna:2024ycx,Paul:2024ujx, Islam:2021mha,Setyawati:2021gom,Wang:2023ueg,Islam:2024zqo, Islam:2024bza,Hinderer:2017jcs,Cao:2017ndf,Liu:2019jpg,Liu:2021pkr,Liu:2023dgl,Ramos-Buades:2021adz}.
Based on the \ac{EOB} formalism \citep{Buonanno:1998gg, Buonanno:2000ef}, two families of eccentric \ac{IMR} waveform templates have been developed, the state of the art of which are \texttt{TEOBResumS-Dal\'i} \citep{Nagar:2024dzj} and \texttt{SEOBNRv5EHM} \citep{Gamboa:2024imd, Gamboa:2024hli}, respectively. 

We employ \texttt{SEOBNRv5EHM}, a time-domain, eccentric, multipolar \ac{BBH} waveform model for binaries with spin aligned or antialigned with the orbital angular momentum.
It incorporates analytical results to the third post-Newtonian order in the \ac{EOB} radiation reaction forces and the waveform modes.
In the zero eccentricity limit, the model reduces to the highly accurate quasi-circular model \texttt{SEOBNRv5HM} \citep{Pompili:2023tna}.
A comparison with numerical relativity waveforms from the Simulating eXtreme Simulations collaboration \citep{Boyle:2019kee,Scheel:2025jct} has shown a high accuracy, with the dominant $(2,\pm 2)$ mode's unfaithfulness (defined in \cref{eq:faithfulness}) being less than 1\%, with a median of 0.02\%, in the total mass range [20, 200] $M_\odot$ up to eccentricity 0.5 at 20 Hz.
This marks an order of magnitude improvement in accuracy compared to the previous generation model \texttt{SEOBNRv4EHM} \citep{Ramos-Buades:2021adz}.

\begin{figure}[htbp]
\begin{center}
\includegraphics[width=\columnwidth]{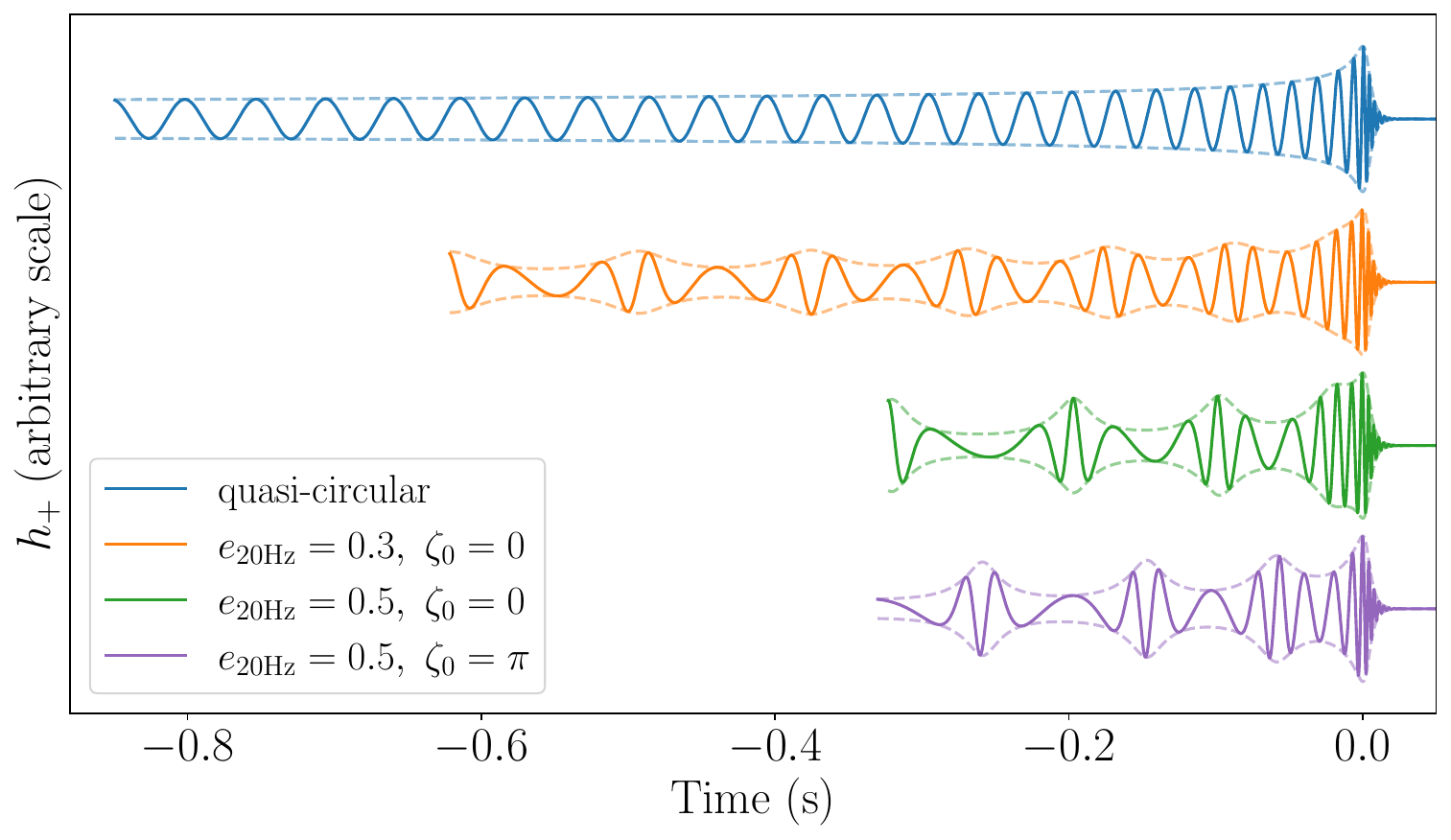}
\caption{This plot showcases the morphology of the \ac{GW} plus polarization from a 30-30 \msun, zero spin \ac{BBH} system generated by \texttt{SEOBNRv5E}. 
The initial eccentricity at 20 Hz is chose to be 0 (labeled as ``quasi-circular"), 0.3 and 0.5 respectively.
For \etwo = 0.5, two cases of the initial relativistic anomaly with $\zeta= 0$ or $\pi$ are presented.
The dashed lines present the amplitude of the combined polarization $h_+ - ih_\times$.
}
\label{fig:waveform}
\end{center}
\end{figure}

The \ac{GW} plus polarization is plotted in \cref{fig:waveform} to showcase the morphology of an eccentric \ac{BBH} system with component masses of 30-30 \msun{} and zero spin.
We consider the initial eccentricity specified at a orbit-averaged \ac{GW} frequency at 20 Hz, \etwo{}, to be $0$, which reduces to the quasi-circular case, and $0.3, 0.5$ for the eccentric scenario.
As a result of a higher \ac{GW} emission power for higher eccentricity, the waveform gets shorter.
For $\etwo=0.5$, we also compare the relativistic anomaly, $\zeta$, to be $0$ or $\pi$, which corresponds to the periastron and apastron for the initial position and a peak or a trough of the \ac{GW} amplitude, respectively.

We use the dominant $(2,\pm2)$ mode of \texttt{SEOBNRv5EHM} (referred to as \texttt{SEOBNRv5E} hereafter) as a template bank to perform a matched-filtering search for eccentric \ac{BBH}.
\red{We ignore higher-order modes to limit the search within available computational resources; incorporating them in the future would be a way to further enhance the search sensitivity.}
As a result, the search space includes six intrinsic parameters, component masses $m_{1/2}$,  component spins $s_{1z/2z}$ aligned with the direction of the total angular momentum, eccentricity \etwo{}, and relativistic anomaly $\zeta$.
All other extrinsic parameters are analytically maximized when matched filtering.

\begin{figure}[htbp]
\begin{center}
\includegraphics[width=\columnwidth]{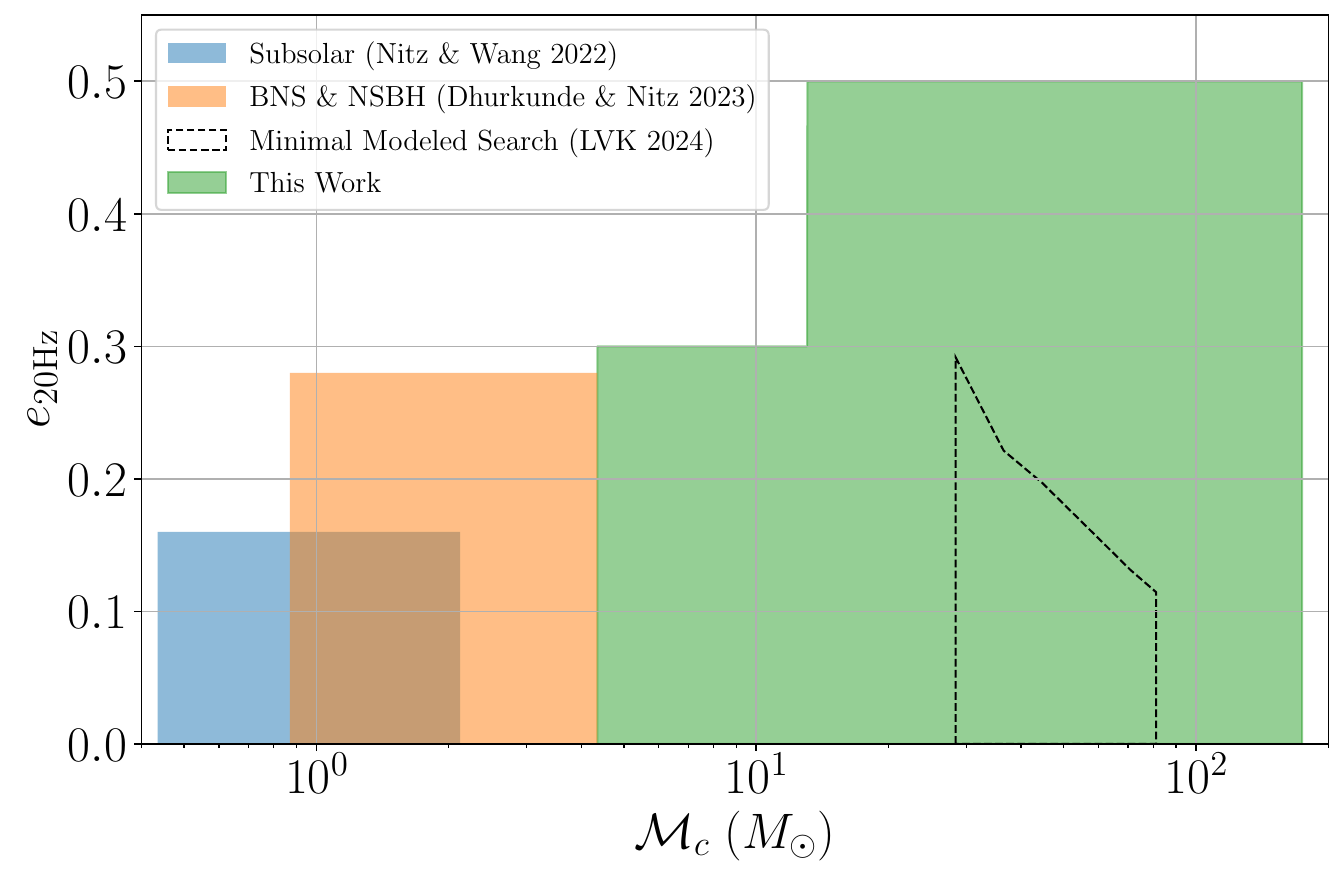}
\caption{The parameter space of the template bank of this work.
In comparison, we also plot the search region of Ref.~\citep{Dhurkunde:2023qoe} for binaries with neutron stars and \citep{Nitz:2022ltl} for binaries with at least one subsolar mass component.
We also plot a region from the minimally modeled search by LVK without using a template \citep{LIGOScientific:2023lpe}.
Note that this region is where the sensitivity is reported by Ref.~\citep{LIGOScientific:2023lpe}.
}
\label{fig:parameter}
\end{center}
\end{figure}

We use a brute-force stochastic placing algorithm \citep{Harry:2009ea,Ajith:2012mn,Kacanja:2024pjh} to construct a bank of discrete templates to cover the targeting parameter space, as the source properties of an astrophysical \ac{GW} signal are unknown a priori.
The component masses are placed in $[5, 200]~\msun$.
The eccentricity \etwo{} is chosen to be up to 0.5 for both component masses greater than 15 \msun, and up to 0.3 otherwise to limit the number of templates given our computational resources.
\red{This choice is made also because eccentric \acp{BBH} are generally expected to be more massive if from hierarchical dynamical formation, as predicted by astrophysical formation channels \citep{Rodriguez:2018pss}.}
The amplitude of the aligned-spin is in the range of [-0.5, 0.5].
Altogether, this choice results in a bank with $\sim860,000$ templates.
A search parameter space is depicted in \cref{fig:parameter} in comparison with the search of subsolar mass binaries \citep{Nitz:2021mzz}, binary neutron stars and neutron star black holes ~\citep{Dhurkunde:2023qoe}, and the minimally modeled search \citep{LIGOScientific:2023lpe}.

A metric to measure the distance in the \ac{GW} signal space is needed when constructing a template bank.
The inner product between two \ac{GW} signals, $\tilde h_1(f)$ and $\tilde h_2(f)$, is defined as
\begin{equation}
    \left(h_1 | h_2 \right) = 4 \Re \int \frac{\tilde{h}_1(f)\tilde{h}^\dagger_2(f)}{S_n(f)} \mathrm{d} f,
\label{eq:innerproduct}
\end{equation}
where $S_n(f)$ is the noise \ac{PSD} and the dagger denotes complex conjugation.
A normalized inner product, or fitting factor \citep{Apostolatos:1995pj}, characterizes the similarity of \ac{GW} signals
\begin{equation}
    \mathcal{O}(h_1, h_2) = \max_{t_c, \phi_c} \frac{\left(h_1|h_2\right)}{\sqrt{\left(h_1|h_1\right)} \sqrt{\left(h_2|h_2\right)}},
\label{eq:faithfulness}
\end{equation}
where $t_c$ and $\phi_c$ are the coalescence time and phase being maximized over.
When constructing the template bank, we request the minimum fitting factor between any \ac{GW} signals in the search region and existing templates in the bank to be 95\%.
Our stochastic placement achieves 94.7\%.

\begin{figure}[htbp]
\begin{center}
\includegraphics[width=\columnwidth]{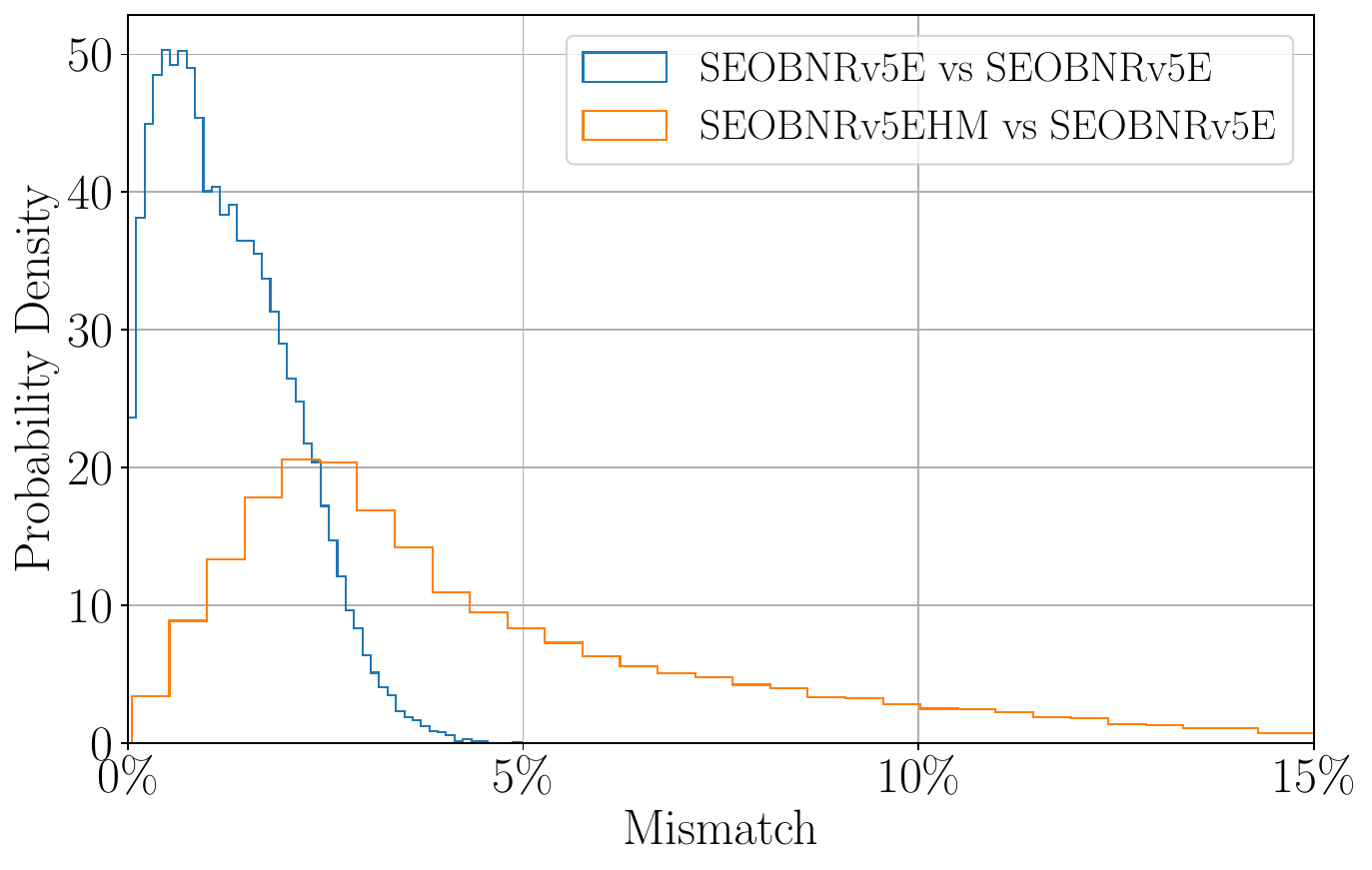}
\caption{The mismatch, which is 1 - fitting factor, comparison between an injection of simulations generated using \texttt{SEOBNRv5E} with only $(2,\pm2)$ mode. 
We also present the results by using all available higher modes with $(2,\pm2), (2,\pm1), (3,\pm3), (3,\pm2), (4,\pm4$), and $(4,\pm3)$ with \texttt{SEOBNRv5EHM}. 
}
\label{fig:ff}
\end{center}
\end{figure}

More specifically, we verify the effectiveness of our template bank by injecting a set of simulations within the search region.
We inject with both \texttt{SEOBNRv5E} with only the dominant modes, and the \texttt{SEOBNRv5EHM} with all available higher modes to characterize the loss of \ac{SNR} against realistic astrophysical signals with higher modes.
The result is shown in \cref{fig:ff}. 
It is shown that our template bank is capable of recovering the \ac{SNR} within a loss of 5\%.
The mean average of the fitting factor is 98.7\%, with the minimum being 94.7\%.
For an injection with higher modes, the mean average drops to 95.4\%, with a minimum being 76.2\%.
Assuming a uniform volumetric distribution of the astrophysical sources, a fitting factor of $x$ translates to a fraction of $1-x^3$ signals to be missed compared with an optimal scenario.
\red{In the future, adding higher-order modes could be a promising development to further increase the search sensitivity.}

\section{Search strategy}
\label{sec:strategy}
We employ the open-sourced, modular, Python package \pycbc \citep{pycbc-github} to search for eccentric \acp{BBH} from LVK's publicly released data of the third observation run (O3) \citep{KAGRA:2023pio}.
O3 observation starts on 2019 April 1, and ends on 2020 March 27, with a one-month break on 2019 October for commissioning and upgrades.
The Advanced LIGO Hanford and Livingston, and Advanced Virgo detectors frequently reached a sky-averaged horizon distance of 110 Mpc, 140 Mpc, and 60 Mpc, respectively, for a fiducial 1.4-1.4 \msun{} binary neutron star source.
In total, O3 observation contains $152$ days of triple-detector observing time, $201.5$ days when both LIGO observatories were observing and 272.1 days when at least two observatories were observing.

The search configuration is similar to the 4th open-gravitational wave catalog (4-OGC) \citep{Nitz:2021zwj}, with the template bank replaced to incorporate eccentric orbital effects and targeted optimization.
Considering only the $(2,\pm2)$ mode, the measured \ac{GW} strain is a projection of the plus and cross polarizations $\tilde{h}_{+/\times} (f)$ 
\begin{equation}
    \tilde{h}(f) = F_+ \tilde{h}_+(f) + F_\times \tilde{h}_\times (f) = A \tilde{h}_+(f) e^{i \Phi},
\end{equation}
where $F_{+/\times}$ is the antenna pattern function.
Replacing the real inner product in \cref{eq:innerproduct} to be complex, and denoting it to $(,)_c$, the maximum matched-filtering \ac{SNR} marginalized over the unknown amplitude $A$ and phase $\Phi$ is 
\begin{equation}
\label{eq:snr}
    \rho_\mathrm{mf} = \frac{\left|(d|h_+)_c \right|}{\sqrt{(h_+|h_+)}}.
\end{equation}
The \ac{GW} strain data is correlated with templates to obtain $\rho_\mathrm{mf}$ time series, and the local maxima are recorded for each observatory.
To distinguish the astrophysical signals from the remaining instrumental or environmental transient noise after data cleaning, $\chi^2$ signal consistency tests are performed by decomposing data in the frequency domain \citep{Allen:2004gu}, denoted as $\chi^2_r$, or in the time-frequency domain, denoted as $\chi^2_{sg}$ \citep{Nitz:2017lco}.
Also, the noise \ac{PSD} variation $v_s$ in the time scale of around $\mathcal{O}(100)~\mathrm{s}$ is recorded to measure the nonstationarity of the background \citep{Zackay:2019kkv, Mozzon:2020gwa}.
Altogether, these quantities are combined with $\rho_\mathrm{mf}$ to construct a statistic, $\rho$, as a likelihood of \ac{GW} in a single observatory
\begin{equation}
    \rho = \frac{\rho_\mathrm{mf}}
    {
    \left[
    \frac{1}{2}(1+(\chi_\mathrm{r}^2)^3)
    \right]^{1/6}
    \left[\frac{1}{4}\chi_\mathrm{r,sg}^2\right]^{1/2}
    v_s^{1/3}
    }.
\end{equation}
Temporal coincident candidates triggered by the same template from two or more detectors are combined and assigned a ranking statistic, representing the likelihood of being genuine \ac{GW} signals through their coherence with astrophysical sources, against empirically measured noise characteristics.
\begin{equation}
    r = \ln\left( \frac{\sigma_\mathrm{min}^3/\bar{\sigma}_\mathrm{HL}^3{}}{A_d \prod_d r_{d}(\rho)} \frac{p(\theta | S)}{p(\theta | N)}\right).
\end{equation}
where $p(\theta|S)/p(\theta|N)$ characterizes the likelihood ratio of the amplitude, phase, and time delay distribution of \ac{SNR} from a signal hypothesis versus the noise, $\sigma_\mathrm{min}^3/\bar\sigma_\mathrm{HL}^3$ characterizes the horizon volume for a template, and $r_d$ is an empirically fitted rate for noise triggers with a ranking $\rho$ in a single detector $d$, and $A_d$ represents the volume of noise time window.
For more details on the derivation of the ranking statistic, refer to, e.g., Ref.~\citep{Davies:2020tsx}.
Last, the \ac{FAR} for search candidates is determined by applying a non-physical time shift exceeding 10 ms for a single detector against another detector. 
This ensures any coincident triggers after time shifting are from background noise, thus the frequency of noise fluctuation mimicking an actual foreground trigger yields the \ac{FAR}.

We perform targeted optimization on the noise rate $r_d$ by smoothing it over templates with similar chirp mass, mass ratio, effective spin and eccentricity.
The smoothing length is empirically tuned until the tail of the ranking statistic from the background noise coincident triggers is suppressed.
We also employ a novel technique to veto potential environmental or instrumental glitches utilizing the O3 glitch catalog from \texttt{Gravity Spy} \citep{glanzer_2021_5649212, Zevin:2016qwy, Zevin:2023rmt}, which is a convolutional neural network to identify and classify the excess power noise into different classes of glitches.
We selected a subset of glitches with high confidence, with \texttt{ml\_confidence} given by the catalog greater than 0.999.
\red{This choice was made by the observation that all \ac{GW} events from 4-OGC \citep{Nitz:2021zwj} would not be misclassified.}

\section{Results}
\label{sec:results}
Throughout the O3 observation, we detect 28 significant \ac{BBH} events with a \ac{FAR} less than once every 100 yr, \red{a convention adopted by the community to claim detections for previously unseen sources \citep{LIGOScientific:2016dsl}.}
All of them are previously identified in both GWTC-3 \citep{KAGRA:2021vkt} and 4-OGC \citep{Nitz:2021zwj}. 
Nevertheless, due to a different template bank employed in our search, we have alternative interpretations that account for eccentricity for all detections.
In addition, 16 \ac{GW} candidates with \acp{FAR} between once per 1 and 100 yr are also reported, all but one of these candidates are also detected as confident events by 4-OGC \citep{Nitz:2021zwj}.
In \cref{tab:significant}, we present the details of all 44 candidates, including their \acp{FAR} and maximum likelihood parameters such as component mass, component spin, effective spin, and eccentricity, together with their matched-filtering \ac{SNR} from each observatory.

\begin{table*}[h]
  \centering
  \caption{This table shows 44 event candidates found by the search with an inverse \ac{FAR} less than 1 per 1 year. Together we show the parameters of the search template that give the maximum ranking statistic for the candidates, this includes component masses $m_1$ and $m_2$, spin amplitudes along the angular momentum direction $s_{1z}$ and $s_{2z}$, the effective spin $\chi_\mathrm{eff}$, and eccentricity $e_\mathrm{20Hz}$. We also report the matched-filtering \ac{SNR} in LIGO Hanford  ($\rho_\mathrm{H}$), LIGO Livingson ($\rho_\mathrm{L}$), and Virgo ($\rho_\mathrm{V}$). }
  \label{tab:significant}
  \begin{tabular}{ c  c  c  c  c  c  c  c  c  c  c  c }
\hline
     & Event & IFAR [yr] & $m_1$ [\msun] & $m_2$[\msun] & $s_{1z}$ & $s_{2z}$ & $\chi_\mathrm{eff}$ & $e_\mathrm{20Hz}$ & $\rho_\mathrm{H}$ & $\rho_\mathrm{L}$ & $\rho_\mathrm{V}$ \\
\hline
1 & GW190408\_181802 & 7279.1 & 29.9 & 20.9 & -0.09 & -0.49 & -0.25 & 0.12 & 9.6 & 9.9 & - \\
2 & GW190412\_053044 & 7279.1 & 50.3 & 6.6 & 0.41 & 0.30 & 0.40 & 0.19 & 8.5 & 14.9 & - \\
3 & GW190413\_134308 & 26.5 & 82.7 & 26.4 & 0.06 & -0.43 & -0.06 & 0.18 & 5.6 & 7.8 & - \\
4 & GW190421\_213856 & 19.2 & 54.0 & 38.1 & -0.40 & -0.26 & -0.34 & 0.22 & 7.4 & 6.6 & - \\
5 & GW190503\_185404 & 2722.3 & 57.1 & 54.3 & 0.34 & 0.37 & 0.35 & 0.08 & 8.9 & 7.5 & - \\
6 & GW190512\_180714 & 206.2 & 24.4 & 17.9 & -0.12 & -0.12 & -0.12 & 0.05 & 5.7 & 10.6 & - \\
7 & GW190513\_205428 & 151.1 & 57.2 & 16.4 & -0.02 & 0.01 & -0.01 & 0.11 & 8.1 & 8.3 & - \\
8 & GW190517\_055101 & 55.7 & 41.9 & 34.1 & 0.42 & 0.39 & 0.41 & 0.14 & 6.4 & 7.6 & - \\
9 & GW190519\_153544 & 9604.8 & 90.0 & 86.3 & 0.47 & 0.45 & 0.46 & 0.19 & 7.9 & 10.5 & - \\
10 & GW190521\_030229 & 7638.5 & 158.3 & 117.3 & 0.28 & 0.25 & 0.27 & 0.15 & 8.6 & 12.3 & - \\
11 & GW190521\_074359 & 54150.3 & 61.1 & 21.1 & -0.27 & 0.11 & -0.17 & 0.03 & 11.9 & 20.4 & - \\
12 & GW190602\_175927 & 10462.6 & 130.2 & 45.8 & 0.48 & -0.01 & 0.35 & 0.05 & 6.3 & 10.7 & - \\
13 & GW190701\_203306 & 1.5 & 64.4 & 57.0 & -0.04 & 0.07 & 0.01 & 0.31 & 5.5 & 10.2 & - \\
14 & GW190706\_222641 & 19.8 & 112.0 & 47.7 & 0.42 & 0.46 & 0.43 & 0.33 & 9.3 & 8.9 & - \\
15 & GW190707\_093326 & 77155.0 & 29.6 & 5.1 & 0.42 & 0.25 & 0.40 & 0.08 & 7.6 & 10.2 & - \\
16 & GW190720\_000836 & 22.9 & 30.0 & 5.5 & 0.49 & 0.46 & 0.49 & 0.04 & 7.2 & 8.0 & - \\
17 & GW190727\_060333 & 8604.8 & 57.1 & 54.3 & 0.34 & 0.37 & 0.35 & 0.08 & 8.4 & 8.0 & - \\
18 & GW190728\_064510 & 8604.8 & 24.5 & 5.9 & 0.45 & -0.28 & 0.31 & 0.20 & 8.7 & 10.5 & - \\
19 & GW190803\_022701 & 2.5 & 56.0 & 43.2 & 0.12 & 0.11 & 0.12 & 0.15 & 5.8 & 6.5 & - \\
20 & 190805\_112555 & 2.3 & 69.0 & 15.8 & 0.50 & 0.22 & 0.45 & 0.47 & 5.4 & 7.5 & - \\
21 & GW190828\_063405 & 8427.9 & 56.6 & 34.2 & 0.39 & 0.44 & 0.41 & 0.02 & 10.4 & 11.6 & - \\
22 & GW190828\_065509 & 133.8 & 33.8 & 11.2 & 0.13 & -0.44 & -0.01 & 0.08 & 7.2 & 7.4 & - \\
23 & GW190915\_235702 & 10727.0 & 46.3 & 21.8 & -0.09 & -0.20 & -0.12 & 0.32 & 10.0 & 7.5 & - \\
24 & GW190924\_021846 & 6890.5 & 10.5 & 5.1 & -0.30 & 0.48 & -0.05 & 0.12 & 6.1 & 11.1 & - \\
25 & GW190925\_232845 & 62.1 & 35.7 & 13.3 & 0.18 & 0.26 & 0.20 & 0.01 & 7.9 & - & 5.2 \\
26 & GW190929\_012149 & 5.1 & 80.0 & 57.8 & -0.07 & 0.11 & 0.00 & 0.29 & 5.7 & 6.9 & - \\
27 & GW190930\_133541 & 130.6 & 26.3 & 5.4 & 0.43 & 0.45 & 0.43 & 0.14 & 6.7 & 8.0 & - \\
28 & GW191105\_143521 & 3.2 & 25.5 & 5.1 & 0.33 & -0.12 & 0.26 & 0.14 & 6.0 & 7.6 & - \\
29 & GW191109\_010717 & 18263.9 & 63.9 & 51.0 & -0.49 & -0.21 & -0.37 & 0.15 & 8.9 & 12.6 & - \\
30 & GW191129\_134029 & 18076.8 & 9.8 & 9.2 & -0.41 & 0.27 & -0.08 & 0.14 & 8.6 & 9.6 & - \\
31 & GW191204\_171526 & 26390.3 & 11.5 & 10.3 & -0.13 & 0.28 & 0.06 & 0.18 & 10.1 & 13.6 & - \\
32 & GW191215\_223052 & 117.1 & 38.7 & 19.4 & -0.13 & 0.36 & 0.03 & 0.25 & 7.1 & 7.7 & - \\
33 & GW191216\_213338 & 28218.5 & 16.1 & 6.8 & 0.27 & 0.04 & 0.20 & 0.14 & 17.3 & - & 5.1 \\
34 & GW191222\_033537 & 6072.7 & 81.5 & 58.8 & -0.13 & 0.48 & 0.12 & 0.16 & 8.0 & 8.1 & - \\
35 & GW191230\_180458 & 28.2 & 77.7 & 67.0 & 0.24 & 0.29 & 0.26 & 0.12 & 7.7 & 6.6 & - \\
36 & GW200128\_022011 & 864.3 & 51.7 & 39.1 & -0.42 & -0.37 & -0.40 & 0.03 & 6.7 & 6.9 & - \\
37 & GW200129\_065458 & 44147.1 & 38.0 & 35.3 & -0.07 & 0.36 & 0.13 & 0.16 & 14.3 & - & 6.8 \\
38 & GW200208\_130117 & 52.2 & 74.1 & 48.6 & 0.25 & 0.37 & 0.30 & 0.17 & 6.3 & 6.5 & - \\
39 & GW200209\_085452 & 1.4 & 57.6 & 31.9 & -0.01 & 0.23 & 0.07 & 0.24 & 7.0 & 6.0 & - \\
40 & GW200219\_094415 & 9.6 & 54.5 & 38.9 & -0.18 & -0.32 & -0.24 & 0.05 & 5.8 & 8.0 & - \\
41 & GW200224\_222234 & 10457.6 & 58.9 & 45.9 & 0.15 & 0.24 & 0.19 & 0.02 & 12.1 & 12.9 & - \\
42 & GW200225\_060421 & 69348.7 & 21.2 & 18.6 & -0.06 & -0.39 & -0.22 & 0.12 & 9.3 & 8.0 & - \\
43 & GW200311\_115853 & 3068.6 & 47.8 & 28.7 & 0.02 & 0.47 & 0.19 & 0.17 & 11.8 & 9.9 & 6.8 \\
44 & GW200316\_215756 & 1.2 & 27.4 & 6.1 & 0.42 & 0.25 & 0.39 & 0.02 & 5.5 & 7.8 & - \\
  \hline
  \end{tabular}
\end{table*}

The subthreshold candidate 190805\_112555 is a discovery not previously documented.
With a \ac{FAR} of just one occurrence every 2.3 yr over an observational period of approximately 200 days with both LIGO observatories, it remains indistinguishable from a random noise fluctuation.
The triggered template is consistent with an eccentric \ac{BBH} with component masses of 69 \msun{} and 15.8 \msun.
It also has the most extreme eccentricity, $e_\mathrm{20Hz} =0.47$, in our catalog.
We examine its constant-Q time-frequency transformation and identify a possible glitch $\sim 0.25$ s prior to the reported merger time.
More details to examine the data quality around this candidate are presented in appendix \ref{sec:dataquality}.

Two out of the three eccentric \ac{GW} events identified by Ref.~\citep{Gupte:2024jfe} by the standard of log$_{10}$ greater than 1, namely GW200129 and GW190701 are both listed in the \cref{tab:significant}.
GW200129 is detected with a high significance, but GW190701 only has a \ac{FAR} once per 1.5 yr.
The chirp mass of the best matched template is 32 \msun{}  and 53 \msun{}, and \etwo{} is 0.16 and 0.31, respectively.

The \ac{BBH} candidate GW200208\_222617 was reported to have non-negligible eccentricity \citep{Romero-Shaw:2022xko, Gupte:2024jfe, Romero-Shaw:2025vbc, McMillin:2025hof}.
Our search finds its \ac{FAR} only once per 0.02 yr due to its low \ac{SNR}, which are 5.8 and 6.2 in LIGO Hanford and Livingston.
It triggers a template with chirp mass of 25 \msun{} and \etwo{} of 0.28.
This event is only detected by the PyCBC-BBH configuration with an astrophysical probability $70\%$ in GWTC-3.
Also note that the astrophysical probability is obtained using a source population inferred by parameter estimation with quasi-circular templates and thus may not be directly applicable for an eccentric population.
With its low significance, we can not distinguish it from noise.

\section{Astrophysical event rate}
\label{sec:astro}
With a conservative assumption that none of the detection of significant \ac{BBH} events are eccentric, we estimate the upper limit of the astrophysical event rates of \ac{BBH} in eccentric orbits.
We use Monte Carlo simulations to robustly characterize the sensitive volume and time, $\langle VT \rangle$, of our search pipeline against a population of simulated astrophysical eccentric \ac{BBH} events, defined as
\begin{equation}
    \langle VT \rangle = T \int \frac{1}{1+z}\frac{\mathrm{d}V}{\mathrm{d}z} f(\theta;z) \mathrm{d}z,
\end{equation}
where $\mathrm{d}V/\mathrm{d}z$ is the differential volume with respect to redshift, $f(\theta; z)$ is the fraction for detectable sources with properties $\theta$ at redshift $z$.
The detection threshold is chosen to be  \ac{FAR} less than once per 100 yr.

We simulate a population of \acp{BBH} generated by \texttt{SEOBNRv5EHM} with higher modes for realistic astrophysical sources and inject them into the LIGO and Virgo O3 data stream.
The exact same algorithm described in section \ref{sec:strategy} is used to search for the injections.
The source parameters $\theta$ are chosen to be discrete values; we consider equal mass binaries and six component masses, \{5,10,15,20,25,30\} \msun, and six eccentricity values $\{0, 0.1, 0.2, 0.3, 0.4, 0.5\}$.
When the component masses are below 15 \msun{}, the eccentricity is limited up to 0.3.
The sky locations and orientations are isotropically distributed.
In total, we inject $10^5$ waveforms into the data with an observation time $\sim0.67$ yr.

\begin{figure}[htbp]
\begin{center}

\includegraphics[width=\columnwidth]{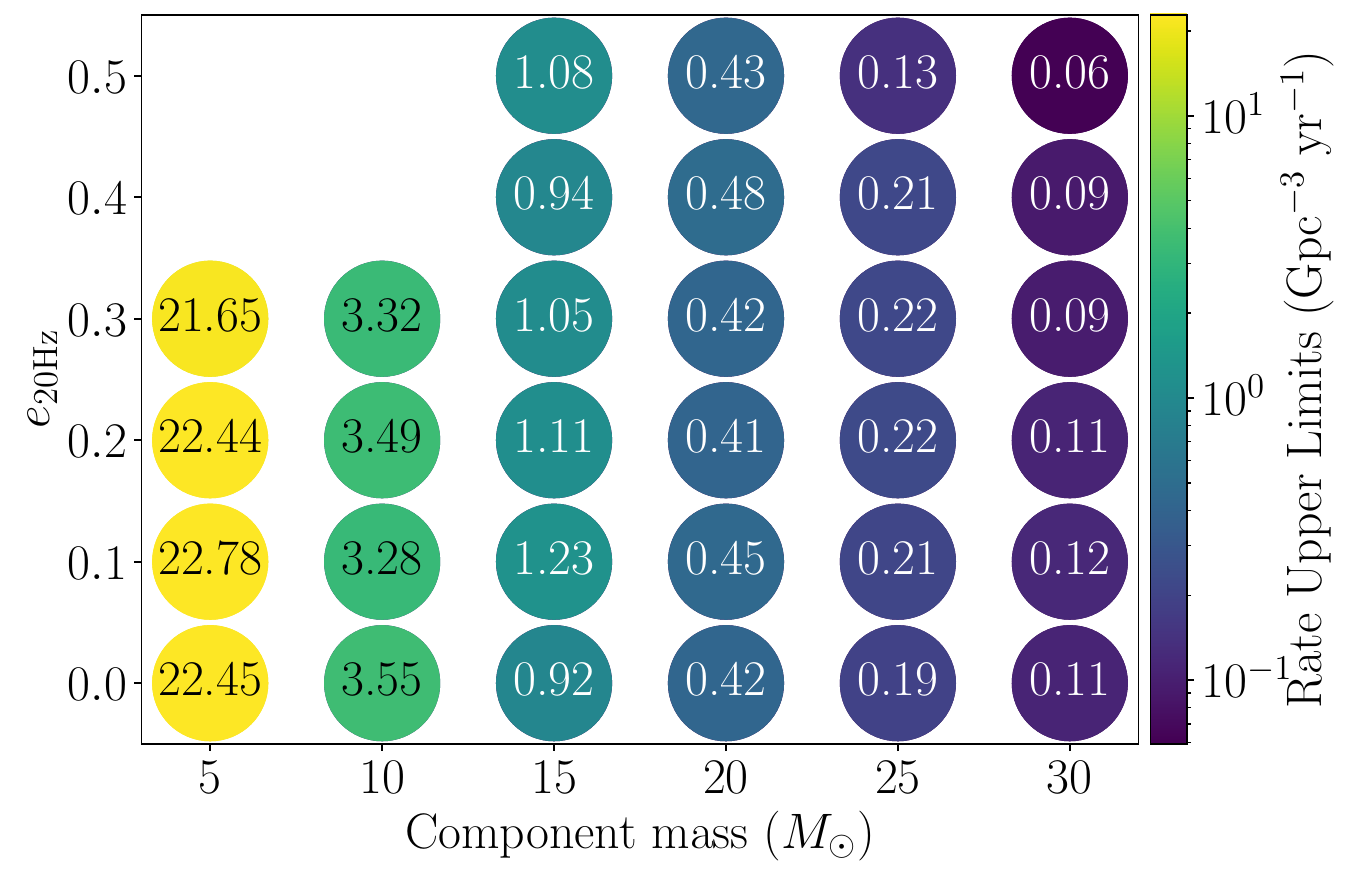}
\caption{The 90\% upper limit for the event rate density, $R_\mathrm{90}$, for a population of eccentric \acp{BBH} with discrete source parameters.
}
\label{fig:rate}
\end{center}
\end{figure}

Using the loudest event statistic \citep{Biswas:2007ni}, which describes the detection as a Poisson process and assumes non-detections, the 90\% upper limit of the event rate density $R_{90}$ is given by $2.3/\langle VT \rangle$.
\Cref{fig:rate} shows our results on $R_\mathrm{90}$.
For a 30-30 \msun{} \ac{BBH} with eccentricity 0.5, we constrain the event rate to be $< 0.06$ Gpc$^{-1}$ yr$^{-1}$.
The sensitive distance corresponding to the 30-30 \msun{} is 2.4 Gpc.
This result is directly comparable with the sensitive distance from the LVK O3 minimal modeled search \citep{LIGOScientific:2023lpe}, which obtains 1.3 Gpc for a total mass 70 \msun{} source.
An approximate comparison between the sensitive distances suggests that our search improves by a factor of 2 for $\sim$ 30 \msun{} eccentric \acp{BBH}, or equivalently, the sensitive volume by a factor of 8.

Our results mark the most stringent event rate upper limits for eccentric \ac{BBH} in the mass range 5 to 30 \msun{}.
\red{Note that these results are only for eccentric \acp{BBH} with discrete source parameters as representative types of sources.}
To convert to any realistic astrophysical distribution, the results should be reweighted against the corresponding mass and eccentricity distribution.
We release our results to enable such reweighting. 

\section{Conclusion}
\label{sec:conclusion}
This work presents the first matched-filtering search for \ac{GW} from eccentric \acp{BBH} in LIGO-Virgo O3 data using an effective-one-body waveform model \texttt{SEOBNRv5E}, demonstrating the feasibility of a matched-filtering \ac{GW} search that incorporates eccentricity in the template bank.
We detect 28 significant events with \ac{FAR} less than 1 per 100 yr, and report an additional 16 candidates with \ac{FAR} ranging from 1 per 100 yr to 1 per 1 yr.
In particular, these include two \ac{GW} events, GW200129 and GW190701, which are claimed to have non-negligible evidence for eccentricity from parameter estimation \citep{Gupte:2024jfe}.
Additionally, all but one candidates has been identified before in the GWTC-3 \citep{KAGRA:2021vkt} and 4-OGC \citep{Nitz:2021zwj} catalogs.
The newly identified candidates cannot be clearly distinguished from noise fluctuations.

With a conservative assumption that no conclusive evidence exists for non-negligible eccentricity in the current \acp{BBH} catalog, we place upper limits on astrophysical event rates for mass in the range 5 to 30 \msun{} and eccentricity up to 0.5.
These provide the most stringent constraints on the potential formation channels for astrophysical eccentric \ac{BBH}, including dynamical formation in  \ac{AGN} disks or globular clusters.

We do not consider single-detector triggers as they require extrapolating the empirically measured noise characteristics. Similarly, we do not consider the astrophysical probability for the candidates, as this requires a well-understood astrophysical prior for the source population.
As our knowledge of such populations evolves, this search can be revisited to unveil weaker signals.
\red{In the future, possible pathways to further increase the search sensitivity include designing better signal consistency tests tailored to eccentric \acp{BBH}, and incorporating higher order modes in the search templates.}

Since May of 2023, LVK has started the fourth observation run with both LIGO observatories approaching a sky-averaged horizon distance of $\sim$160 Mpc and Virgo for $\sim$60 Mpc \footnote{https://observing.docs.ligo.org/plan/} \citep{KAGRA:2013rdx}.
The \ac{GW} astronomy has entered an era of routine detections with the frequency of a few events every week.
In addition, the next generation ground-based \ac{GW} observatories, such as Einstein Telescope \citep{punturo2010third,Hild:2010id,abernathy2011einstein} and Cosmic Explorer \citep{LIGOScientific:2016wof,cewhitepaper, Evans:2021gyd}, are projected with significant sensitivity improvement in the low frequency band from 2 to 10 Hz, where the eccentricity is more prominent than in the higher frequency band.
The ability to search and characterize eccentric binaries would become increasingly important.

The search results and all scripts necessary to reproduce this work are released in \url{https://github.com/gwastro/O3-eccentricBBH-search}.

\acknowledgements
The computational work for this manuscript was carried out on the Hypatia computer cluster at the Max Planck Institute for Gravitational Physics (Albert Einstein Institute) in Potsdam.
Y.-F.W. thanks Aldo Gamboa for significant support in using \texttt{SEOBNRv5EHM}, Nihar Gupte, Antoni Ramos Buades, Hector Estelles, 
Lorenzo Pompili, Alessandra Buonnano for many useful discussions and feedback, and Steffen Grunewald, Raffi Enficiaud for technical support in computation..
We acknowledge the Syracuse University HTC Campus Grid and NSF award ACI-1341006

This research has made use of data or software obtained from the Gravitational Wave Open Science Center (gwosc.org), a service of the LIGO Scientific Collaboration, the Virgo Collaboration, and KAGRA. This material is based upon work supported by NSF's LIGO Laboratory which is a major facility fully funded by the National Science Foundation, as well as the Science and Technology Facilities Council (STFC) of the United Kingdom, the Max-Planck-Society (MPS), and the State of Niedersachsen/Germany for support of the construction of Advanced LIGO and construction and operation of the GEO600 detector. Additional support for Advanced LIGO was provided by the Australian Research Council. Virgo is funded, through the European Gravitational Observatory (EGO), by the French Centre National de Recherche Scientifique (CNRS), the Italian Istituto Nazionale di Fisica Nucleare (INFN) and the Dutch Nikhef, with contributions by institutions from Belgium, Germany, Greece, Hungary, Ireland, Japan, Monaco, Poland, Portugal, Spain. KAGRA is supported by Ministry of Education, Culture, Sports, Science and Technology (MEXT), Japan Society for the Promotion of Science (JSPS) in Japan; National Research Foundation (NRF) and Ministry of Science and ICT (MSIT) in Korea; Academia Sinica (AS) and National Science and Technology Council (NSTC) in Taiwan.

\appendix
\section{Data quality checks around 190805\_112555}
\label{sec:dataquality}

\begin{figure*}[h]
\begin{center}
\includegraphics[width=\columnwidth]{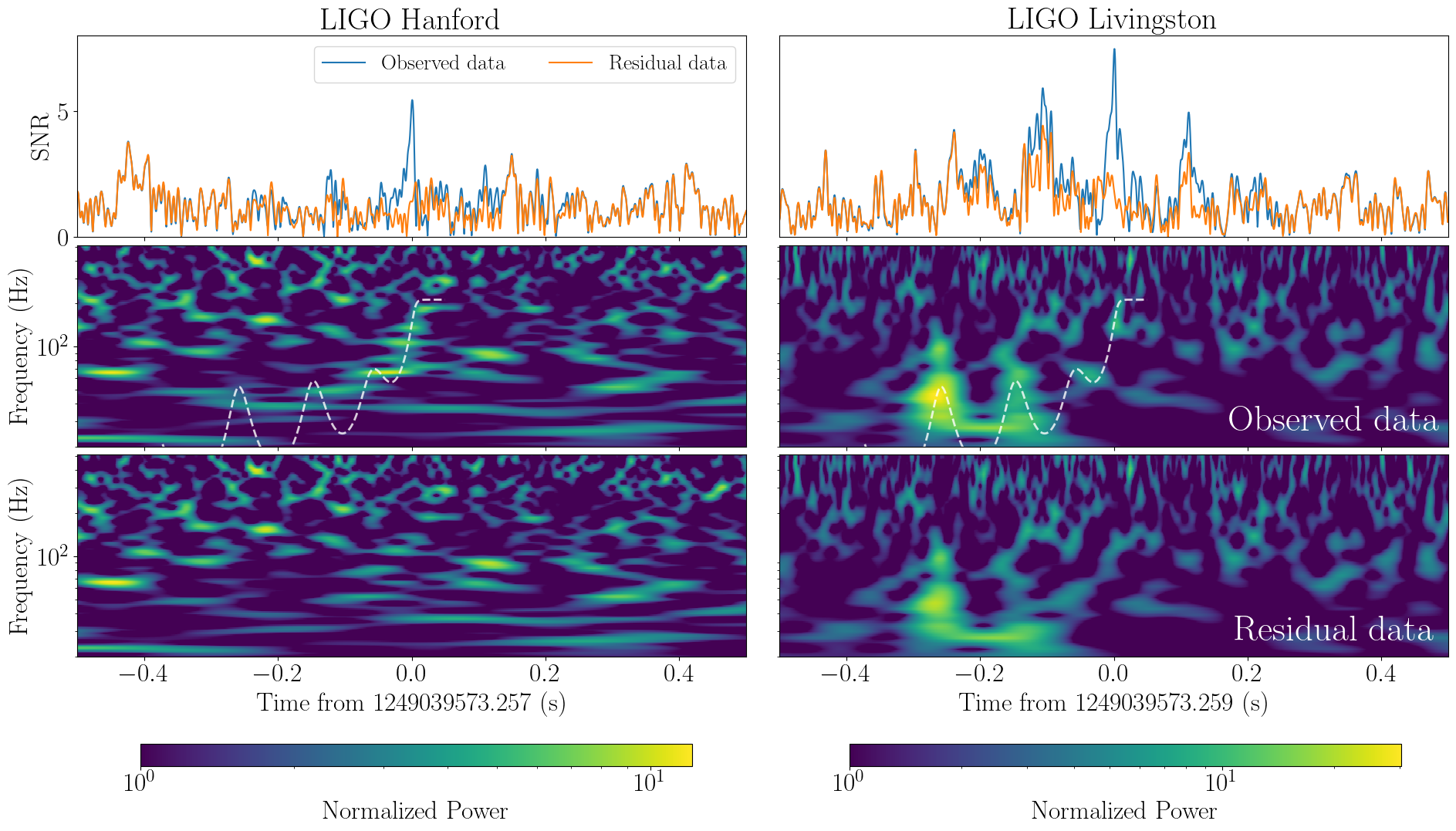}
\caption{In the first row, we plot the matched-filtering \ac{SNR} (defined in \cref{eq:snr} with the observed data from LIGO Hanford and LIGO Livingston, and the residual data subtracting the best matched templates from matched-filtering.
The \ac{SNR} time series peak no longer exists in the residual data.
The middle rows represent the constant-Q short-time Fourier transformation of the observed data, overplotted by white dashed lines for the frequency-time evolution of the best-matched templates, which is consistent with an eccentric \ac{BBH} with component masses 69 and 15.8 \msun, component spin $s_{1z}$ and $s_{2z}$ being 0.5 and 0.22, respectively, and an eccentricity at 20 Hz of 0.47.
The last row presents the constant-Q transformation with the residual data by subtracting the best-matched template.
}
\label{qscan}
\end{center}
\end{figure*}

We perform an in-depth follow-up examination on the data quality around the candidate 190805\_112555.
The results are shown in \cref{qscan}.
From a constant-Q short-time Fourier transform, a glitch around 0.25 s prior to the merger time in LIGO Livingston is identified.
Without a glitch model, we can not isolate the contribution to the \ac{SNR} from the glitch.
Subtracting the best-matched \ac{GW} template from the data, LIGO Hanford data is apparently stable and Gaussian; however, the excess power remains in LIGO Livingston.
This can also be seen in the \ac{SNR} time series. After subtracting the template, LIGO Livingston still has an \ac{SNR} peak at $\sim$4 around 0.1 s before the identified merger time, while the \ac{SNR} time series peak no longer exists in LIGO Hanford.

We conclude from our examination that this trigger is likely to originate from the contamination of a glitch in LIGO Livingston.

\bibliography{ref.bib}

\end{CJK*}
\end{document}